\documentclass[prd,aps,twocolumn,preprintnumbers, showpacs, nofootinbib,superscriptaddress,notitlepage]{revtex4-1}
\usepackage{amssymb,amsthm,amsmath}
\usepackage{textcomp}
\usepackage{subfig,graphicx}   
\usepackage{color}      
\usepackage{slashed}    
\usepackage{verbatim}
\usepackage[normalem]{ulem}
\usepackage{rotating}   
\usepackage{multirow}   
\begin{document}
\title{ SU(3) symmetry analysis in charmed baryon two body decays with penguin diagram contribution
}

\affiliation{ Department of Physics and Institute of Theoretical Physics, Nanjing Normal University, Nanjing, Jiangsu 210023, China}
\affiliation{ School of Physics, East China University of Science and Technology, Shanghai 200237, China}
\affiliation{ Shanghai Key Laboratory of Particle Physics and Cosmology, School of Physics and Astronomy, Shanghai Jiao Tong University,
Shanghai 200240, China}
\affiliation{Particle Theory and Cosmology Group, Center for Theoretical Physics of the Universe, Institute for Basic Science (IBS), Daejeon 34126, Korea }
\affiliation{School of Materials Science and Physics, China University of Mining and Technology, Xuzhou 221116, China}

\author{Zhi-Peng Xing$^{1}$}
\email{zpxing@nnu.edu.cn}
\author{Yu-Ji Shi$^{2,3}$}
\email{shiyuji@ecust.edu.cn}
\author{Jin Sun$^{4}$}
\email{sunjin0810@ibs.re.kr(corresponding author)}
\author{Ye Xing$^{5}$}
\email{xingye_guang@cumt.edu.cn}

\preprint{CTPU-PTC-24-20}

\begin{abstract}

An increasing number of experimental measurements from the BESIII, Belle, and Belle-II collaborations encourage investigations into  charmed baryon two-body decay processes.
By including contributions from the penguin diagrams that are ignored in previous studies, we perform a global analysis with SU(3) flavor symmetry.
Assuming all form factors are real, we achieve a remarkable minimal $\chi^{2}/d.o.f = 0.788$ and find that the contribution of the amplitude proportional to $V_{cb}^*V_{ub}$ is of the order $\sim 0.01$, comparable with the contribution of the tree-level diagram.
Additionally, by using the KPW theorem to reduce the number of amplitudes from 13 to 7 in the leading contribution, it becomes possible to consider the complex form factor case for the leading IRA amplitude in the global analysis.
However, the analysis of complex form factors significantly conflicts with the experimental data $Br(\Xi_c^0\to\Xi^-\pi^+)$, and by excluding this data, $\chi^2/d.o.f$ is reduced from 5.95 to 1.19.
Although the analysis of complex form factors shows a significant central value of the penguin diagram contribution, the large errors from the corresponding form factors make it a challenge to precisely determine its true contribution.
Consequently, the direct CP violation in decay processes is predicted to be approximately zero.
With more data in future experiments, the penguin diagram contribution with the amplitude proportional to $V_{cb}^*V_{ub}$ will be precisely determined, allowing for a more accurate prediction of CP violation.
Our analysis necessitates further theoretical investigations and experimental measurements in the future.

\end{abstract}

\maketitle

\section{Introduction}

In the past decade, the study of baryon decays has attracted increasing attention in particle physics, especially in heavy flavor physics.
With the development of experimental facilities like LHC, SuperKEKB, and BEPC, an increasing number of baryon decay channels and new baryon states are being measured for the first time~\cite{Belle:2013htj,LHCb:2015yax,BESIII:2015bjk,BESIII:2015ysy,LHCb:2017iph,LHCb:2017uwr,Belle:2017ext,BESIII:2019nep,LHCb:2024lao,Belle:2024cmc}. 
A large amount of experimental data has greatly promoted corresponding theoretical research since it provides a new platform to precisely test the Standard Model (SM).
Recently, there have been many interesting and influential works on the characters~\cite{Liu:2022igi,Deng:2023csv,Su:2024lzy,Han:2023xbl,Liu:2023pwr,Deng:2023qaf,Tan:2023opd,Han:2024ucv,Peng:2024pyl}, decay behaviors~\cite{Li:2017ndo,Zhang:2021oja,Geng:2022osc,Miao:2022bga,Aliev:2022tvs,Rui:2023fiz,Wang:2024oyi,Shen:2024aba} and angular distributions of baryon~\cite{Li:2022nim,Fu:2023ose}.

With the emergence of increasing theoretical and experimental research, one of the most important issues in particle physics, CP violation (CPV), has been proposed.
The study of CPV not only helps in precisely testing the Standard Model (SM) but also promotes a deeper understanding that may lead to the discovery of new sources of CPV explaining matter-antimatter asymmetry.
Therefore, the measurement of CPV in baryon decays is significant, and many experimental collaborations attempt to detect CPV in baryon decay~\cite{LHCb:2017hwf,Belle:2022uod,BESIII:2023drj,BESIII:2024nif}. 
Unfortunately, there is no conclusive evidence showing nonzero CPV in baryon decays so far.
Given the current stage where experimental data are inadequate for effectively observing CPV in all baryon decay processes, theoretical guidance to identify channels with significant CPV is crucial.
In pursuit of this strategy, there have been many theoretical works aiming to predict CP violation in baryons~\cite{Bediaga:2020qxg,Zhang:2022iye,Zhang:2022emj,Yelton:2022hai,Shen:2023eln,Du:2024jfc}.

Previous studies of charmed baryon two-body decay (CBTD) processes  indicated that CPV in CBTD is typically  at the order of $O(10^{-4})$~\cite{Sakharov:1967dj}.
 For many years such small CPV is rarely mentioned despite the abundance of experimental data from collaborations such as BESIII, Belle, Belle-II, and LHCb.
However, recent studies have indicated that CBTD processes may indeed exhibit observable CPV~\cite{He:2024pxh,Sun:2024mmk}.
As indicated in our previous work~\cite{Sun:2024mmk}, the potential source of CPV largely originates from the penguin diagram involving the CKM matrix element $\lambda_b=V_{cb}^*V_{ub}$, aligning with traditional understanding of CPV.
To confirm this assumption and assess whether the penguin diagram can indeed produce observable CPV in CBTD processes, it is crucial and necessary to estimate its contribution.

In the analysis of CBTD, SU(3) symmetry serves as a powerful tool that has been extensively utilized in the study of heavy hadrons~\cite{Lu:2016ogy,Zhao:2018mov,Geng:2019xbo,Jia:2019zxi,Wang:2020gmn,Huang:2021aqu,Xing:2023dni,Geng:2023pkr,Zhong:2024qqs,Wang:2024ztg}. 
In the absence of detailed dynamic understanding, the effectiveness of the imposed symmetry can be assessed by how well the known data fit, as indicated by a good $\chi^2$ per degree of freedom ($\chi^2/d.o.f$).
Including the experimental measurement in this year, the number of experimental data is increased to 35 shown in Tables.~\ref{data} and \ref{dataeta}.  This allows us to revisit the global analysis of CBTD, incorporating the contribution from the penguin diagram.

The layout of this article is given as follows. In Sec. II, we give the theoretical framework of CBTD including the penguin diagram contribution in IRA method of SU(3) analysis. 
In Sec. III, we give the global analysis of CBTD for real form factors by considering the added penguin SU(3) amplitudes. 
In Sec. IV, we extend to the complex form factors case and further predict the direct CPV 
$A_{CP}$ of CBTD and give some discussion. 
The last part draws the conclusion.

\section{ Theoretical framework including the penguin operator in SU(3) analysis}\label{se_su3}

For  CBTD processes, the effective weak interaction Hamiltonian is 
\begin{eqnarray}\label{lagrangian}
\mathcal{H}_{eff}&=&\frac{G_F}{\sqrt{2}}\bigg(\sum_{\lambda,i=1,2}C_i \lambda O_i-\sum_{j=3}^6 C_j \lambda_b O_j\bigg)+h.c.,
\end{eqnarray}
where $O_{1,2}$ is the current-current operator and $O_{3\sim 6}$ is QCD penguin operator. $\lambda=V^*_{cq}V_{uq'}$ with $q,q'=d,s$ 
and $\lambda_b=V^*_{cb}V_{ub}$. 

For the current-current operator, the specific expressions are
\begin{eqnarray}
\sum_\lambda\lambda O_i&=&(V^*_{cs}V_{ud}O^{s\bar d u}_i+V^*_{cq}V_{uq}O^{q \bar qu}+V^*_{cd}V_{us}O_i^{d\bar s u})\notag\\
 O^{q_1\bar q_2 q_3}_1&=&(\bar{q}_{1\alpha} c_\beta)_{V-A}(\bar{q_3}_\beta {q_2}_{\alpha})_{V-A},\notag\\
 O^{q_1\bar q_2 q_3}_2&=&(\bar{q}_{1\alpha} c_\alpha)_{V-A}(\bar{q_3}_\beta {q_2}_{\beta})_{V-A},q=d,s.\label{H}
 \end{eqnarray}
By considering the flavor of quark, the matrix of  Hamiltonian $H^{ij}_k$ can be defined as 
\begin{eqnarray}
\sum_\lambda\lambda O_1&=&H^{ij}_k\times (\bar{q}_{i\alpha} c_\beta)_{V-A}(\bar{q_j}_\beta q^k_{\alpha})_{V-A},\notag\\
\sum_\lambda\lambda O_2&=&H^{ij}_k \times(\bar{q}_{i\alpha} c_\alpha)_{V-A}(\bar{q_j}_\beta q^k_{\beta})_{V-A},
 \end{eqnarray}
 where $\{i,j,k\}=\{1,2,3\}$ with $\{q_{1},q_2,q_3\}=\{u,d,s\}$. One  finds the matrix of Hamiltonian only depends on the flavor of operators. Therefore we can give a general SU(3)  irreducible decomposition in irreducible representation amplitude (IRA) method.  The Hamiltonian   is decomposed as:  $3\otimes \bar 3\otimes 3=3\oplus 3\oplus\bar 6\oplus 15$ with the relations as
 \begin{eqnarray}
H^{ij}_k&=&\frac{1}{2}(H_{15})^{ij}_k-\frac{1}{2}(H_{\bar 6})^{ij}_k-\frac{1}{8}(H_3)^{ij}_k+\frac{3}{8}(H_{3^\prime})^{ij}_k,\notag\\
(H_{15})^{ij}_k&=&-\frac{1}{4}(H^{im}_m\delta^j_k+H^{jm}_m\delta^i_k+H^{mi}_m\delta^j_k+H^{mj}_m\delta^i_k)\notag\\
&&+H^{ij}_k+H^{ji}_k,\notag\\
(H_{\bar 6})^{ij}_k&=&\frac{1}{2}(H^{im}_m\delta^j_k-H^{jm}_m\delta^i_k-H^{mi}_m\delta^j_k+H^{mj}_m\delta^i_k)\notag\\
&&-H^{ij}_k+H^{ji}_k,\notag\\
(H_3)^{ij}_k&=&H^{mi}_m\delta^j_k+H^{jm}_m\delta^i_k,\notag\\(H_{3\prime})^{ij}_k&=&H^{im}_m\delta^j_k+H^{mj}_m\delta^i_k.\label{irah}
 \end{eqnarray}
 It is noted that in previous work, the decomposition has been presented in an alternative form~\cite{He:2000ys, Wang:2020gmn}.
 We demonstrate in Appendix A that the two decomposition forms are equivalent. 

Based on the aforementioned decomposition relationship, we can derive the Hamiltonian matrix in IRA corresponding to each operator in Eq.~\ref{H}.
 For the Cabibbo-allowed operator induced by  $c\to s\bar d u$, the IRA Hamiltonian are
 \begin{eqnarray}
(H_{\bar 6})^{31}_2=- (H_{\bar 6})^{13}_2=(H_{15})^{31}_2= (H_{15})^{13}_2=V_{cs}^*V_{ud},
 \end{eqnarray}
 while, for doubly Cabibbo-suppressed  transition induced by $c\to d  \bar s u $, we have
 \begin{eqnarray}
 (H_{\bar 6})^{21}_3=- (H_{\bar 6})^{12}_3=(H_{15})^{21}_3= (H_{15})^{12}_3=V_{cd}^*V_{us}.
 \end{eqnarray}
For Cabibbo-suppressed transition induced by  $c\to u\bar d d$, we have
 \begin{eqnarray}
(H_3)^1&=&\lambda_d,\notag\\
(H_{\bar 6})^{21}_2&=&-(H_{\bar 6})^{12}_2=(H_{\bar 6})^{13}_3=-(H_{\bar 6})^{31}_3=\frac{1}{2}\lambda_d,\notag\\
\frac{1}{3}(H_{15})^{21}_2&=&\frac{1}{3}(H_{15})^{12}_2=-\frac{1}{2}(H_{15})^{11}_1\notag\\
&=&-(H_{15})^{13}_3=-(H_{15})^{31}_3=\frac{1}{4}\lambda_d,
 \end{eqnarray}
 where $\lambda_d=V_{cd}^*V_{ud}$ and $(H_3)^i=(H_{3})^{ji}_k\delta^k_j$. Meanwhile, for the Cabibbo-suppressed transition by $c\to u\bar s s$, we have 
 \begin{eqnarray}
(H_3)^1&=&\lambda_s,\quad \notag\\
(H_{\bar 6})^{12}_2&=&-(H_{\bar 6})^{21}_2=(H_{\bar 6})^{31}_3=-(H_{\bar 6})^{13}_3=\frac{1}{2}\lambda_s,\notag\\
\frac{1}{3}(H_{15})^{31}_3&=&\frac{1}{3}(H_{15})^{13}_3=-\frac{1}{2}(H_{15})^{11}_1\notag\\
&=&-(H_{15})^{12}_2=-(H_{15})^{21}_2=\frac{1}{4}\lambda_s,
 \end{eqnarray}
 where $\lambda_s=V_{cs}^*V_{us}$.
By combining all Hamiltonian matrices corresponding to the operators, one can directly infer the relation $\lambda_b = V_{cb}^* V_{ub}$, which satisfies the condition $\lambda_b + \lambda_d + \lambda_s = 0$.
Therefore,  the Hamiltonian matrix for Cabibbo-suppressed processes by $c\to ud\bar d/s\bar s$ are expressed as
\begin{eqnarray}
&&(H_{\bar 6})^{31}_3=-(H_{\bar 6})^{13}_3=(H_{\bar 6})^{12}_2=-(H_{\bar 6})^{21}_2=\frac{1}{2}(\lambda_s-\lambda_d),\notag\\
&& (H_{15})^{31}_3=(H_{15})^{13}_3=\frac{3}{4}\lambda_s-\frac{1}{4}\lambda_d=\frac{\lambda_s-\lambda_d}{2}-\frac{\lambda_b}{4},\notag\\
&&(H_{15})^{21}_2=(H_{15})^{12}_2=\frac{3}{4}\lambda_d-\frac{1}{4}\lambda_s=\frac{\lambda_d-\lambda_s}{2}-\frac{\lambda_b}{4},\notag\\
&&(H_{15})^{11}_1=\frac{\lambda_b}{2},(H_3)^1=-\lambda_b.\label{iralambdab}
\end{eqnarray}
We found that the terms $(H_{15})^{11}_1$ and $(H_{3})^1$ are both proportional to $\lambda_b$, introducing a new weak phase. When the operators $c \to ud\bar{d}/s\bar{s}$  contribute to CBTD processes together, this new phase from $\lambda_b$ emerges naturally, particularly in final state interaction scenarios where nearly all Cabibbo-suppressed processes can introduce $\lambda_b$~\cite{He:2024pxh}. Traditionally, terms proportional to $\lambda_b$ are neglected due to its small value, $\lambda_b \sim O(10^{-4}) \ll \lambda_{d,s}$. However, to incorporate the contribution of the penguin operator, terms proportional to $\lambda_b$ in $O_{1,2}$ must be included.

In fact, $\lambda_b$ is not only introduced by $O_{1,2}$, but also  by the penguin operators
\begin{eqnarray}
 O_{3,5}&=&(\bar{u}_\alpha c_\alpha)_{V-A}\sum_{q=u,d,s}(\bar{q}_\beta {q}_{\beta})_{V\mp A},\notag\\
 O_{4,6}&=&(\bar{u}_\beta c_\alpha)_{V-A}\sum_{q=u,d,s}(\bar{q}_\alpha {q}_{\beta})_{V\mp A},
\end{eqnarray}
Similarly,  applying the operator to the IRA decomposition relation in Eq.~\ref{irah}, one can derive $(H_3)^1 = \lambda_b$.
It is found that both current-current operator $O_{1,2}$ and the penguin operator $O_{3\sim 5}$ contribute to $H_3$. 
This makes it difficult to distinguish the sources of $H_3$ principally. 
Therefore, in our work, we will only focus on estimating the contribution of $H_3$ without considering its source.

 One can find that the both  $O_{1,2}$ and penguin operator  contribute to the $H_3$ and in principle the souces of $H_3$ can not be distinguished.  Therefore, in our work, we will only estimate the contribution of the $H_3$ without distinguish its source.

Besides the above Hamiltonian in CBTD, the initial and final states of the charmed baryon decay processes can be expressed by $3\times 3$ matrix  as
\begin{eqnarray}
T_{c\bar3}&=&
\begin{pmatrix}
0& \Lambda_c^+ &\Xi_c^+ \\
-\Lambda_c^+ & 0&\Xi_c^0\\
-\Xi_c^+& -\Xi_c^0&0
\end{pmatrix},
P=
\begin{pmatrix}
\frac{\pi^0+\eta_q}{\sqrt{2}}& \pi^+ &K^+ \\
\pi^- & \frac{-\pi^0+\eta_q}{\sqrt{2}}&K^0\\
K^-& \bar{K}^0&\eta_s
\end{pmatrix},\notag\\
T_8&=&
\begin{pmatrix}
\frac{\Sigma^0}{\sqrt{2}}+\frac{\Lambda^0}{\sqrt{6}}& \Sigma^+ &p \\
\Sigma^- & -\frac{\Sigma^0}{\sqrt{2}}+\frac{\Lambda^0}{\sqrt{6}}&n\\
\Xi^-& \Xi^0&-\frac{2\Lambda^0}{\sqrt{6}}
\end{pmatrix}.
\end{eqnarray}
Here the anti-triplet charmed baryon can also be expressed as $(T_{c\bar3})_i=\epsilon^{ijk}(T_{c\bar 3})^{[jk]}=(\Xi_c^0,-\Xi_c^+,\Lambda^+_c)$, and the $\eta_s$ and $\eta_q$ are the mixture of $\eta_1$ and $\eta_8$: $\eta_8=\eta_q/\sqrt{3}-\eta_s\sqrt{2}/\sqrt{3},\quad \eta_1=\eta_q \sqrt{2}/\sqrt{3}+\eta_s/\sqrt{3}$.
To analyze the experimental data, we  consider the physical mixing effects with  
 $\eta^{(\prime)}$  states 
\begin{eqnarray}
&&\begin{pmatrix}
\eta \\
\eta^\prime\\
\end{pmatrix}=\begin{pmatrix}
\cos\phi&-\sin\phi \\
\sin\phi&\cos\phi\\
\end{pmatrix}\begin{pmatrix}
\eta_q \\
\eta_s\\
\end{pmatrix},
\end{eqnarray}
where  the physical mixing angles  $\phi=(39.3\pm1.0)^{\circ}$~\cite{Gan:2020aco}.

By  defining $(H_{\bar 6})_{ij}=\frac{1}{2}\epsilon_{kmi}(H_{\bar 6})^{km}_j$, we contract the flavor indices to obtain all the SU(3) invariant decay amplitudes as~\cite{Huang:2021aqu}
\begin{eqnarray}
\mathcal{M}^{IRA}
&=&a_{15} \times(T_{c\bar{3}})_i(H_{15})^{\{ik\}}_j(\overline{T_8})^j_kP^l_l\notag\\
&+&b_{15} \times(T_{c\bar{3}})_i(H_{15})^{\{ik\}}_j(\overline{T_8})^l_kP^j_l\notag\\
&+&c_{15} \times(T_{c\bar{3}})_i(H_{15})^{\{ik\}}_j(\overline{T_8})^j_lP^l_k\notag\\
&+&d_{15} \times(T_{c\bar{3}})_i(H_{15})^{\{jk\}}_l(\overline{T_8})^l_jP^i_k\notag\\
&+&e_{15} \times(T_{c\bar{3}})_i(H_{15})^{\{jk\}}_l(\overline{T_8})^i_j P^l_k\notag\\
&+&a_{6} \times(T_{c\bar{3}})^{[ik]}(H_{\overline{6}})_{\{ij\}}
(\overline{T_8})^j_kP^l_l\notag\\
&+&b_{6} \times(T_{c\bar{3}})^{[ik]}(H_{\overline{6}})_{\{ij\}}(\overline{T_8})^l_kP^j_l\notag\\
&+&c_{6} \times(T_{c\bar{3}})^{[ik]}(H_{\overline{6}})_{\{ij\}}(\overline{T_8})^j_lP^l_k\notag\\&+&d_{6} \times(T_{c\bar{3}})^{[lk]}(H_{\overline{6}})_{\{ij\}}(\overline{T_8})^i_kP^j_l.\notag\\
&+&a_{3} \times(T_{c\bar{3}})_{i}(H_{3})^{k}(\overline{T_8})^i_l P^l_k.\notag\\
&+&b_{3} \times(T_{c\bar{3}})_{i}(H_{3})^{k}(\overline{T_8})^i_k P^l_l.\notag\\
&+&c_{3} \times(T_{c\bar{3}})_{i}(H_{3})^{i}(\overline{T_8})^k_l P^l_k.\notag\\
&+&d_{3} \times(T_{c\bar{3}})_{i}(H_{3})^{l}(\overline{T_8})^k_l P^i_k.\label{su3}
\end{eqnarray}
Here for different ways of contraction, we assign an undetermined parameter, respectively. 
By absorbing the CKM matrix element $V_{cb}^*V_{ub}$ into SU(3) amplitudes $a_3, b_3, c_3, d_3$, the $H_3$ in the penguin operator can be defined as $(H_3)^1=1$. For the $H_{15}$, since the $\lambda_b$  appears with $\lambda_d-\lambda_s$ together in $(H_{15})^{13}_3,(H_{15})^{31}_3,(H_{15})^{12}_2,(H_{15})^{21}_2$, one can omit the $\lambda_b$ in these matrix elements and set $-\lambda_d=\lambda_s=\sin\theta= 0.2265$.  Then all the amplitudes of anti-triplet charmed baryon two-body decays can be expressed by these SU(3) amplitudes as shown in Tables.~\ref{tableamplitute} and \ref{tableamplituteeta}.

Generally, each of the SU(3) irreducible amplitudes in Eq.~\ref{su3} can be expressed by two form factors: the parity conserving one $f^q$ and the parity violating one $g^q$, which reads as  
\begin{eqnarray}\label{def}
  q_{3,6}&=&G_F\bar{u}(f^q_{3,6} - g^q_{3,6}\gamma_5)u,\quad q=a,b,c,d,\notag\\
  q_{15}&=&G_F\bar{u}(f^q_{15} - g^q_{15}\gamma_5)u,\quad q=a,b,c,d,e.\label{eq:su3amps}
\end{eqnarray}
Expressed by these form factors, the experimental data can be analyzed by a global fit.

 \begin{table*}[htbp!]
\caption{The SU(3) amplitudes  of anti-triplet charmed baryons  decays into an octet baryon and an octet meson.}\label{tableamplitute}
\begin{tabular}{|c|c|c|c|c|c|c|c}\hline\hline
channel &SU(3) amplitude \\\hline \hline
$\Lambda^{+}_{c}\to \Sigma^{0}  \pi^{+} $ & $ (b_{15}-b_6-c_{15}+c_6+d_6)/\sqrt{2}$\\\hline
$\Lambda^{+}_{c}\to \Lambda  \pi^{+} $ & $ -(-b_{15}+2 e_{15}+b_6-c_{15}+c_6+d_6)/\sqrt{6}$\\\hline
$\Lambda^{+}_{c}\to \Sigma^{+}  \pi^{0} $ & $ (-b_{15}+b_6+c_{15}-c_6-d_6)/\sqrt{2}$\\\hline
$\Lambda^{+}_{c}\to p  K_{S}^{0} $ & $ ( \sin^2\theta \left(d_{15}-d_6+e_{15}\right)+b_6-e_{15}-b_{15})/\sqrt{2}$\\\hline
$\Lambda^{+}_{c}\to \Xi^{0}  K^{+} $ & $ -c_6+c_{15}+d_{15}$\\\hline
$\Xi^{+}_{c}\to \Sigma^{+}  K_{S}^{0} $ & $  (\sin^2\theta (-b_{15}+b_{6}-e_{15})+e_{15}-d_6+d_{15})/\sqrt{2}$\\\hline
$\Xi^{+}_{c}\to \Xi^{0}  \pi^{+} $ & $ -d_{15}-d_6-e_{15}$\\\hline
$\Xi^{0}_{c}\to \Sigma^{0}  K_{S}^{0} $ & $  (\sin^2\theta \left(e_{15}-b_6-b_{15}\right)+(c_{15}-e_{15}+c_6+d_6))/2$\\\hline
$\Xi^{0}_{c}\to \Lambda  K^0_S$   &$ \sin^2\theta \left(b_{15}+b_6-2c_{15}-2c_6-2d_6+e_{15}\right)/2\sqrt{3}+ (2b_{15}+2 b_6-c_{15}- c_6-d_6-e_{15})/2\sqrt{3}$
\\\hline
$\Xi^{0}_{c}\to \Sigma^{+}  K^{-} $ & $ c_6+c_{15}+d_{15}$\\\hline
$\Xi^{0}_{c}\to \Xi^{-}  \pi^{+} $ & $ b_6+b_{15}+e_{15}$\\\hline
$\Xi^{0}_{c}\to \Xi^{0}  \pi^{0} $ & $ (-b_6+d_6-b_{15}+d_{15})/\sqrt{2}$\\\hline\hline
 $\Lambda^{+}_{c}\to \Sigma^{0}  K^{+} $   & $ (d_3+\sin\theta \left(b_{15}-b_6+d_{15}+d_6\right))/\sqrt{2}+\lambda_b d_{15}/2\sqrt{2}$\\\hline
$\Lambda^{+}_{c}\to \Lambda  K^{+} $  & $ (-2a_3+d_3-\sin\theta \left(-b_{15}+b_6+2c_{15}-2 c_6+3d_{15}+d_6+2 e_{15}\right))/\sqrt{6}+\lambda_b d_{15}/2\sqrt{6}$\\\hline
$\Lambda^{+}_{c}\to \Sigma^{+} K^0_S/K^0_L $   & $ (d_3+\sin\theta \left(b_{15}-b_6+d_6-d_{15}\right))/\sqrt{2}$\\\hline
$\Lambda^{+}_{c}\to p  \pi^{0} $          & $ (a_3+\sin\theta \left(c_{15}+e_{15}-c_6-d_6\right))/\sqrt{2}+\lambda_b e_{15}/2\sqrt{2}$\\\hline
$\Lambda^{+}_{c}\to n  \pi^{+}$           & $a_3 -\sin\theta \left(-c_{15}+e_{15}+c_6+d_6\right)$\\\hline
$\Xi^{+}_{c}\to \Sigma^{0}  \pi^{+} $     & 
$ (a_3-d_3)/\sqrt{2}-\sin\theta \left(-b_{15}+b_6+c_{15}-c_6+d_{15}+e_{15}\right)/\sqrt{2}-\lambda_b d_{15}/2\sqrt{2}$\\\hline
$\Xi^{+}_{c}\to \Lambda  \pi^{+} $   &
$ (-a_3-d_3)/\sqrt{6}+\sin\theta \left(b_{15}-b_6+c_{15}-c_6+3d_{15}+2d_6+e_{15}\right)/\sqrt{6}-\lambda_b d_{15}/2\sqrt{6}$ \\\hline
$\Xi^{+}_{c}\to \Sigma^{+}  \pi^{0} $    & $ -(a_3-d_3)/\sqrt{2}+\sin\theta \left(-b_{15}+b_6+c_{15}-c_6-d_{15}-e_{15}\right)/\sqrt{2}-\lambda_b e_{15}/2\sqrt{2}$\\\hline
$\Xi^{+}_{c}\to p  K^0_S/K^0_L $          & $ \mp(-d_3+\sin\theta( b_{15}-b_6-d_{15}+d_6))/\sqrt{2}$\\\hline
$\Xi^{+}_{c}\to \Xi^{0}  K^{+} $          & $ -a_3-\sin\theta \left(-c_{15}+e_{15}+c_6+d_6\right)$\\\hline
$\Xi^{0}_{c}\to \Sigma^{0}  \pi^{0} $     & $  (a_3+2c_3+d_3+\sin\theta \left(-b_{15}-b_6-c_{15}-c_6+d_{15}+e_{15}\right))/2+\lambda_b(b_{15}+c_{15}+d_{15}+e_{15})/4$\\\hline
$\Xi^{0}_{c}\to \Lambda  \pi^{0} $   & $ (a_3+d_3+\sin\theta \left(e_{15}+b_{15}+b_6+c_{15}+c_6-3d_{15}-2 d_6\right))/2 \sqrt{3} +\lambda_b(b_{15}+c_{15}+d_{15}+e_{15})/4\sqrt{3}$\\\hline
$\Xi^{0}_{c}\to \Sigma^{+}  \pi^{-} $     & $c_3+d_3 -\sin\theta  \;(c_6+c_{15}+d_{15})+\lambda_b b_{15}/2$\\\hline
$\Xi^{0}_{c}\to p  K^{-} $              & $ c_3+d_3+\sin\theta \;(c_{15}+c_6+d_{15})+\lambda_b b_{15}/2$\\\hline
$\Xi^{0}_{c}\to \Sigma^{-}  \pi^{+} $     & $ a_3+c_3-\sin\theta \left(e_{15}+b_{15}+b_6\right)+\lambda_bc_{15}/2$\\\hline
$\Xi^{0}_{c}\to n  K^0_S/K^0_L $& $ \mp (c_3 +\sin\theta \left(-b_{15}-b_6+c_{15}+c_6+d_6\right))/\sqrt{2}$\\\hline
$\Xi^{0}_{c}\to \Xi^{-}  K^{+} $          & $ a_3+c_3+\sin\theta \left(b_{15}+b_6+e_{15}\right)+\lambda_b c_{15}/2$\\\hline
$\Xi^{0}_{c}\to \Xi^{0}  K^0_S/K^0_L $          & $ (c_3+ \sin\theta \left(b_{15}+b_6-c_{15}-c_6-d_6\right))/\sqrt{2}$\\\hline\hline
$\Lambda^{+}_{c}\to p  K^0_L $ 		& $ ( \sin^2\theta \left(d_{15}-d_6+e_{15}\right)-b_6+e_{15}+b_{15})/\sqrt{2}$\\\hline
$\Lambda^{+}_{c}\to n  K^{+} $ 		& $ \sin^2\theta \left(d_6+d_{15}+e_{15}\right)$\\\hline
$\Xi^{+}_{c}\to \Sigma^{0}  K^{+} $ & 
$\sin^2\theta \left(-b_{15}+b_6+e_{15}\right)/\sqrt{2}$\\\hline
$\Xi^{+}_{c}\to \Lambda  K^{+} $&
$\sin^2\theta \left(-b_{15}+b_6+2c_{15}-2c_6-2d_6-e_{15}\right)/\sqrt{6}$\\\hline
$\Xi^{+}_{c}\to \Sigma^{+}  K^0_L $ & $ ((b_6-b_{15}-e_{15})\sin^2\theta+d_6-d_{15} -e_{15})/\sqrt{2}$\\\hline
$\Xi^{+}_{c}\to p  \pi^{0} $ 		& 
$ ((c_6-c_{15}+d_{15})\sin^2\theta)/\sqrt{2}$\\\hline
$\Xi^{+}_{c}\to n  \pi^{+} $ 		& 
$ (-c_{15}+c_6-d_{15})\sin^2\theta$\\\hline
$\Xi^{0}_{c}\to \Sigma^{0}  K_{L}^{0}$ & $  (\sin^2\theta \left(e_{15}-b_6-b_{15}\right)+(-c_{15}+e_{15}-c_6-d_6))/2$\\\hline
$\Xi^{0}_{c}\to \Lambda  K^0_L$  &$ \sin^2\theta \left(b_{15}+b_6-2c_{15}-2 c_6-2 d_6+e_{15}\right)/2\sqrt{3}+ (-2b_{15}-2 b_6+c_{15}+ c_6+d_6+e_{15})/2\sqrt{3}$\\\hline
$\Xi^{0}_{c}\to p  \pi^{-} $ 		& $ \sin^2\theta \; (c_6+c_{15}+d_{15})$\\\hline
$\Xi^{0}_{c}\to \Sigma^{-}  K^{+} $ & $ \sin^2\theta \left(b_{15}+e_{15}+b_6\right)$\\\hline
$\Xi^{0}_{c}\to n  \pi^{0} $ 		
& $-\sin^2\theta\; (c_6+c_{15}-d_{15})/\sqrt{2}$\\\hline
\hline
\end{tabular}
\end{table*}

\begin{table*}[htbp!]
\caption{The SU(3) amplitudes  of anti-triplet charmed baryons  decays into an octet baryon and  $\eta$ or $\eta^\prime$.}\label{tableamplituteeta}
\begin{tabular}{|c|c|c|c|c|c|c|c}\hline\hline
channel &SU(3) amplitude \\\hline \hline
$\Lambda^{+}_{c}\to \Sigma^{+}  \eta $ & $\cos\phi(2a_{15}-2a_6+b_{15}-b_6+c_{15}-c_6+d_6)/\sqrt{2}+\sin\phi\;(a_6-a_{15}-d_{15})$\\\hline
$\Lambda^{+}_{c}\to \Sigma^{+}  \eta^\prime $ & $\sin\phi(2a_{15}-2 a_6+b_{15}-b_6+c_{15}-c_6+d_6)/\sqrt{2}-\cos\phi\;(a_6-a_{15}-d_{15})$\\\hline
\multirow{3}{*}{$\Lambda^{+}_{c}\to p  \eta $}      & $(\sqrt{2}\cos\phi(a_3+2b_3)-2(b_3+d_3)\sin\phi)/2$\cr
&$+\sin\theta\big(\cos\phi \left(2a_{15}-2 a_6+c_{15}-c_6+d_6-e_{15}\right)/\sqrt{2}+ \sin\phi \left(-a_{15}+a_6-b_{15}+b_6-d_{15}-e_{15}\right)\big)$\cr
&+$\lambda_b\sqrt{2}\cos\phi\; e_{15}/4$\\\hline
\multirow{3}{*}{$\Lambda^{+}_{c}\to p  \eta^\prime $ }      & $(2(b_3+d_3)\cos\phi+\sqrt{2}\sin\phi(a_3+2b_3))/2$\cr
&$+\sin\theta\big(\sin\phi \left(2a_{15}- 2a_6+c_{15}-c_6+d_6-e_{15}\right)/\sqrt{2}+ \cos\phi \left(a_{15}-a_6+b_{15}-b_6+d_{15}+e_{15}\right)\big)$\cr
&+$\lambda_b\sqrt{2}\sin\phi\; e_{15}/4$\\\hline
\multirow{3}{*}{$\Xi^{+}_{c}\to \Sigma^{+} \eta $ } & $b_3\sin\phi-(a_3+2b_3+d_3)\cos\phi/\sqrt{2}$\cr
&$+\sin\theta(\cos\phi(2a_{15}-2a_6+b_{15}-b_6+c_{15}-c_6+d_{15}+e_{15})/\sqrt{2}- \sin\phi (a_{15}-a_6+d_6-e_{15}))$\cr
&-$\lambda_b\sqrt{2}\cos\phi \;e_{15}/4$\\\hline
\multirow{3}{*}{$\Xi^{+}_{c}\to \Sigma^{+}  \eta^\prime $}   &$-b_3 \cos\phi-(a_3+2b_3+d_3)\sin\phi/\sqrt{2}$\cr
&$+\sin\theta \big(\sin\phi(2a_{15}-2a_6+b_{15}-b_6+c_{15}-c_6+d_{15}+e_{15})/\sqrt{2}+\cos\phi(a_{15}-a_6+d_6-e_{15})\big)$\cr
&-$\lambda_b\sqrt{2}\sin\phi \;e_{15}/4$\\\hline
$\Xi^{+}_{c}\to p  \eta $ 		& $ \sin^2\theta (\cos\phi(-2a_{15}+2a_6-c_{15}+c_6-d_{15})/\sqrt{2}-\sin\phi\left(-a_{15}+a_6-b_{15}+b_6-d_6\right))$\\\hline
$\Xi^{+}_{c}\to p  \eta^\prime $ 		& $\sin^2\theta(\sin\phi(-2a_{15}+2a_6-c_{15}+c_6-d_{15})/\sqrt{2}+\cos\phi(-a_{15}+a_6-b_{15}+b_6-d_6))$\\\hline
$\Xi^{0}_{c}\to \Xi^{0}  \eta $ & $ \cos\phi(2a_{15}+2 a_6+b_{15}+b_6+d_{15}-d_6)/\sqrt{2}-\sin\phi(a_{15}+a_6+c_{15}+c_6)$\\\hline
$\Xi^{0}_{c}\to \Xi^{0}  \eta^\prime $ & $ \sin\phi(2a_{15}+2 a_6+b_{15}+b_6+d_{15}-d_6)/\sqrt{2}+\cos\phi(a_{15}+a_6+c_{15}+c_6)$\\\hline
\multirow{3}{*}{$\Xi^{0}_{c}\to \Sigma^{0}  \eta $ }   & $ (\cos\phi(a_3+2b_3+d_3)-\sqrt{2}b_3\sin\phi)/2$\cr
&$+\sin\theta \big(\cos\phi\left(2a_{15}+2 a_6+b_{15}+b_6+c_{15}+c_6+d_{15}-e_{15}\right)/2
-\sin\phi \left(a_{15}+a_6-d_6+e_{15}\right)/\sqrt{2}\big)$\cr
&+$\lambda_b(\cos\phi(2a_{15}+b_{15}+c_{15}+d_{15}+e_{15})-\sqrt{2}\sin\phi\; a_{15})/4$\\\hline
\multirow{3}{*}{$\Xi^{0}_{c}\to \Sigma^{0}  \eta^\prime$ }  & $ (\sqrt{2}b_3\cos\phi+\sin\phi(a_3+2b_3+d_3))/2$\cr
&$+\sin\theta (\sin\phi\left(2a_{15}+2 a_6+b_{15}+b_6+c_{15}+c_6+d_{15}-e_{15}\right)/2+\cos\phi \left(a_{15}+a_6-d_6+e_{15}\right)/\sqrt{2}\big)$\cr
&+$\lambda_b(\sqrt{2}\cos\phi\; a_{15}+\sin\phi(2a_{15}+b_{15}+c_{15}+d_{15}+e_{15}))/4$\\\hline
\multirow{4}{*}{$\Xi^{0}_{c}\to \Lambda  \eta $}  & $(\cos\phi(a_3+2b_3+2c_3+d_3)-\sqrt{2}\sin\phi\; (b_3-2c_3))/2\sqrt{3}$\cr
&$+ \sin\theta \big( -\cos\phi \left(6a_{15}+6 a_6+b_{15}+b_6+c_{15}+c_6+3d_{15}-2 d_6+e_{15}\right)/(2 \sqrt{3})$\cr
&$-\sin\phi\left(-3a_{15}-3 a_6-2b_{15}-2 b_6-2c_{15}-2 c_6+d_6+e_{15}\right)/\sqrt{6}\big)$\cr
&-$\lambda_b(\cos\phi(-2a_{15}-b_{15}-c_{15}-d_{15}-e_{15})+\sqrt{2}\sin\phi\; a_{15})/4\sqrt{3}$\\\hline
\multirow{4}{*}{$\Xi^{0}_{c}\to \Lambda  \eta^\prime$ }  &$(\sqrt{2}\cos\phi \;(b_3-2c_3)+\sin\phi(a_3+2b_3+2c_3+d_3)/2\sqrt{3}$\cr &$+\sin\theta \big( -\sin\phi \left(6a_{15}+6 a_6+b_{15}+b_6+c_{15}+c_6+3d_{15}-2 d_6+e_{15}\right)/(2 \sqrt{3})$\cr
&$+\cos\phi\left(-3a_{15}-3 a_6-2b_{15}-2 b_6-2c_{15}-2 c_6+d_6+e_{15}\right)/\sqrt{6}\big)$\cr
&+$\lambda_b(\sqrt{2}\cos\phi\; a_{15}+\sin\phi(2a_{15}+b_{15}+c_{15}+d_{15}+e_{15}))/4\sqrt{3}$\\\hline
$\Xi^{0}_{c}\to n \eta $        & $\sin^2\theta(\cos\phi\left(2a_{15}+2 a_6+c_{15}+c_6+d_{15}\right)/\sqrt{2}- \sin\phi \left(a_{15}+a_6+b_{15}+b_6-d_6\right)\big)$\\\hline
$\Xi^{0}_{c}\to n \eta^\prime $        & $\sin^2\theta(\sin\phi\left(2a_{15}+2 a_6+c_{15}+c_6+d_{15}\right)/\sqrt{2}
+\cos\phi  \left(a_{15}+a_6+b_{15}+b_6-d_6\right)\big)$\\\hline
\hline
\end{tabular}
\end{table*}

\section{ Global analysis of charmed baryon two-body decays with penguin operator}

Along with a significant amount of experimental data in Tables~\ref{dataeta} and \ref{data}, it is imperative to revisit the global analysis of SU(3) irreducible amplitudes and further determine the contribution from the penguin diagrams.  Since the experimental measurements for branching ratios of $\Xi_c^0$ are obtained in terms of the ratios between $Br(\Xi_c\to B M)$ and $Br(\Xi_c\to \Xi^-\pi^+)$, we use these ratios as data for the fit directly. In Tables~\ref{dataeta} and \ref{data}, it is evident that among the measured processes, only $\Xi_c^0 \to \Xi^- K^+$ depends on the amplitude $c_3$. However, with only one measurement, $Br(\Xi_c^0 \to \Xi^- K^+)/Br(\Xi_c^0 \to \Xi^- \pi^+)=0.0275(57)$, the corresponding two form factors cannot be determined. Therefore, the amplitude $c_3$ cannot be determined by the present data. 
Fortunately,  the KPW theorem~\cite{Korner:1970xq, Pati:1970fg} requires that the color structure of the two-quark operators in the effective Hamiltonians is symmetric, which conflicts with the antisymmetric color structure of quarks in the baryon state.
Since, in $c_3$ term in Eq.~(\ref{su3}), the two-quark fields in the charmed baryon are directly connected to the Hamiltonian $H_3^i$, the contribution of $c_3$ is suppressed by the color symmetry.
Therefore, in our work, we assume $f_3^c = g_3^c = 0$ in Eq.~(\ref{eq:su3amps}).

Generally, the form factors $f_q$ and $g_q$ in Eq.(\ref{eq:su3amps}) are complex numbers.
However, the current experimental data are insufficient for us to analyze the complex form factors,  involving 51 independent parameters.
Therefore, at this stage, we assume that all form factors are real numbers.
Note that since  the CKM matrix element $|\lambda_b|$ has been absorbed into $H_3$ amplitude,   the corresponding form factors   should have weak phase. 
However, since strong phases are absent in the real form factor case, considering the weak phase alone is meaningless.
Based on these form factors, the experimental observables, including  the  branching ratio and polarization parameter $\alpha$,  are expressed as following~\cite{He:2015fsa,Huang:2021aqu}
\begin{eqnarray}
&&\frac{d\Gamma}{d\cos\theta_M}=\frac{G_{F}^{2}|\vec{p}_{B_{n}}|(E_{B_n}+M_{B_{n}})}{8\pi M_{B_c}}(|F|^2+\kappa^2 |G|^2)\notag\\
&&\qquad \qquad \times(1+\alpha \hat\omega_i\cdot\hat p_{B_n} ),\notag\\
&&\alpha=\frac{2\rm{Re}(F*G)\kappa}{(|F|^2+\kappa^2 |G|^2)},\;\;
\beta=\frac{2\mathrm{Im}(F^*G)\kappa}{(|F|^2+\kappa^2 |G|^2)},\notag\\
&&\gamma=\frac{|F|^2-|\kappa G|^2}{(|F|^2+\kappa^2 |G|^2)},\;\;
\kappa=\frac{|\Vec{p}_{B_n}|}{(E_{B_n}+M_{B_n})}.\label{observable}
\end{eqnarray}
where $\hat\omega_i$ and $\hat p_{B_n}$ are the unit vector of initial state spin and final state momentum, respectively.  $F(G)$ are linear functions of $f_i(g_i)$ depending on the specific processes.  

From the amplitudes with the final states $\eta^{(\prime)}$ in Table \ref{tableamplituteeta}, we found that the parameters $a_6$ and $a_{15}$ only exist in terms of the combinations $a_6 \pm a_{15}$. Additionally, the processes depending on $a_6 - a_{15}$ are relatively fewer, which makes the fit results for $a_6$ and $a_{15}$ have much larger uncertainty. Therefore, in order to obtain the precise value of $a_6 + a_{15}$, we redefines the new SU(3) irreducible amplitude and corresponding form factors as
\begin{eqnarray}
&& a=a_{6}-a_{15},\quad a^{\prime}=a_{6}+a_{15},\notag\\
&& f^{a}=f_{6}^{a}-f_{15}^{a},\quad g^{a}=g_{6}^{a}-g_{15}^{a},\notag\\
&&f^{a\prime}=f_{6}^{a}+f_{15}^{a},\quad g^{a\prime}=g_{6}^{a}+g_{15}^{a}.
\end{eqnarray}
To correctly describe the phenomenological results of CBTD processes, we performed a global fit with updated experimental data in Table \ref{data} using the nonlinear least-$\chi^2$ method~\cite{Fit}. 
By assuming real form factors for simplicity, we perform fits for real form factors in Eq.~\ref{su3} and obtain the fit results (Case I) in Table \ref{tablefit} with the corresponding prediction in Table.~\ref{dataeta} and \ref{data}.
The $\chi^2/d.o.f = 0.788$ (d.o.f means degree of freedom) indicates that our fit is reasonable and that the real form factors effectively explain the current experimental data. 
Using the fitted form factors, we can predict other observables of CBTD processes that have not yet been measured in experiments as shown in Tables.~\ref{dataeta} and \ref{data}.

Our predictions reveal that the decay modes involving $\eta$ and $\eta^\prime$ in the final state exhibit different behaviors. 
Taking the $\Lambda_c^+$ decays as an example, the branching ratios of the $\Lambda_c^+ \to \Sigma^+$ modes in both cases are comparable, whereas those of the $\Lambda_c^+ \to p$ modes differ significantly. 
This intriguing phenomenon also manifests in $\Xi_c$ decay modes. 
These distinct behaviors can be attributed to the fitted form factors in Table.~\ref{tablefit}, which show that the largest values are
$g_a=0.126(10)$, $g_{a\prime}=0.160(32)$ and $g_6^b=-0.186(14)$. 
Since $\eta$ and $\eta^\prime$ are mixed states of $\eta_q$ and $\eta_s$, the amplitudes of charmed baryon decay modes involving $\eta$ and $\eta^\prime$ in the final state can be expressed as a combination of the decay amplitudes with $\eta_q$ and $\eta_s$, parameterized by the mixing angle $\phi=(39.3\pm1.0)^{\circ}$, as shown in  Table.~\ref{tableamplituteeta}.
In the case of the $\Lambda_c^+ \to \Sigma^+ \eta_{q/s}$ modes, the cancellation between the form factors $g^a = g_6^a - g_{15}^a$ and $g^6_b$ reduces the $\Lambda_c^+ \to \Sigma^+ \eta_q$ decay amplitude, making the $\Lambda_c^+ \to \Sigma^+ \eta_s$ amplitude the dominant contribution.
Consequently, this leads to comparable branching ratios for the $\Lambda_c^+ \to \Sigma^+$ modes.
In the $\Lambda_c^+ \to p \eta_{q/s}$ modes, the dominant contributions come from the amplitudes $a_3$, $b_3$, and $d_3$, making the $\Lambda_c^+ \to p \eta_{q/s}$ amplitudes comparable.
Consequently, the two modes with $\eta_{q/s}$ interfere constructively and destructively in the amplitude of $\Lambda_c^+ \to p \eta^{(\prime)}$, leading to significantly different branching ratios.

For the polarization parameters $\alpha$, the situation becomes significantly more complex because they reflect the discrepancies between P-conservation and P-violation factors.
Consequently, the predictions we offer tend to have greater uncertainties.
Thus, it is most appropriate to address this topic once sufficient experimental data is available, enabling more accurate results that account for complex phase factors.

The numerical results in Table.~\ref{tablefit} show that the amplitudes corresponding to $H_3$ have  non-zero value. Since the CKM matrix element $V_{cb}^*V_{ub}$ is absorbed into these amplitudes,  the order of these amplitudes are 
\begin{eqnarray}
A_3\sim V_{cb}^*V_{ub}\times \langle B P|\mathcal{H}|B_c\rangle\sim O(10^{-2}).
\end{eqnarray}
This order is at the same level as the matrix element of the current-current operator. It is very interesting to note that for the CKM matrix element with order $O(10^{-4})$, the hadronic matrix element, which contains strong and other long-distance interactions, may be large.

Despite the fit showing a nonzero value for the amplitude with $H_3$, we still cannot completely determine the true contribution of $H_3$. This is because the form factors we used are real, and the complete sets of independent parameters in the IRA (isospin rotation analysis) have not been fully used. In particular, when considering long-distance interactions, it is possible that the form factors contain imaginary parts, making the assumption of real form factors inappropriate~\cite{Cao:2023csx,He:2024pxh}.
 Therefore we expect to completely determine the penguin operator contribution with more experiment measurement data in the future.

 \begin{table*}[htbp!]
\caption{
The global fit results for the real (Case I in Eq.~\ref{su3})  and complex form factors (Case II in Eq.~\ref{su3reduce}).  }
\label{tablefit}
\begin{tabular}
{|c|c|c|c|c|c|c|c|c|c|}\hline\hline
form factors  &\multicolumn{5}{c|}{ Case I ($\chi^{2}$/d.o.f=0.788)}  \cr\cline{1-6}
\multirow{3}{*}{vector(f) } 
&$f^a=-0.0615(34)$ &$f^{b}_{6}=0.0592(32)$ 
 &$f^{c}_{6}= 0.0186(49)$
 & $f^d_{6}=0.0064(36)$ 
 &  
\\\cline{2-6}
& $f^{a\prime}=-0.016(13)$ 
&$f^b_{15}=-0.0089(17)$ 
&$f^c_{15}=-0.0213(23)$ 
& $f^d_{15}=-0.0056(44)$
 & $f^e_{15}=-0.0250(27)$
\\\cline{2-6}
&$f^{a}_{3}=0.0058(10)$
& $f^{b}_{3}=-0.0270(24)$  
& 
& $f^{d}_{3}=0.03076(91)$
& \\\cline{2-6}
\hline
\multirow{3}{*}{axial-vector(g) } 
&$g^a=0.126(10)$ 
&$g^{b}_{6}=-0.186(14)$ 
& $g^{c}_{6}= -0.0437(61)$
 & $g^d_{6}=0.037(12)$ 
 &  \\\cline{2-6}
&$g^{a\prime}=0.160(32)$ 
& $g_{15}^{b}=0.056(11)$ 
 & $g^c_{15}=-0.043(18)$
& $g^d_{15}=0.006(16)$
& $g^e_{15}=-0.104(12)$
\\\cline{2-6}
&$g^{a}_{3}=0.0215(28)$
& $g^{b}_{3}=0.0210(94)$ 
& 
& $g^d_{3}=-0.0818(52)$
&\\\cline{2-6}
\hline\hline
\multirow{2}{*}{form factors } &\multicolumn{5}{c|}{ Case II ($\chi^{2}$/d.o.f=1.19)}  \cr\cline{2-6}&\multicolumn{3}{c|}{absolute value} & \multicolumn{2}{c|}{strong phase}\\\cline{1-6}
\multirow{3}{*}{vector(f) } 
&$f^a_6=0.022(23)$ 
&$f^{b}_{6}=0.023(51)$ 
 &$f^{c}_{6}= 0.0415(87)$
 &$\delta f^b_6=-1.088(813)$ 
 & $\delta f^c_6=0.951(729)$ 
\\\cline{2-6}
& $f^d_{6}=-0.018(31)$ 
&  
& $f^e_{15}=0.026(41)$
& $\delta f^d_{6}=-2.593(735)$ 
 & $\delta f^e_{15}=-0.534(682)$ 
\\\cline{2-6}
&$f^{a}_{3}=0.0024(55)$
& $f^{b}_{3}=0.01(14.29)$ 
&
& $\delta f^{a}_{3}=1.695(811)$ 
&  $\delta f^{b}_{3}=-3.042(1.371)$ 
 \\\cline{2-6}
\hline
\multirow{4}{*}{axial-vector(g) } 
&$g^a_6=0.184(67)$ 
&$g^{b}_{6}=0.29(11)$ 
& 
 & $\delta g^a_6=1.268(817)$ 
 & $\delta g^b_6=-2.624(0.829)$
 \\\cline{2-6}
&  $g^{c}_{6}= 0.014(54)$
& $g^d_{6}=0.062(99)$
 & 
& $\delta g^{c}_{6}=0.942(655)$
& $\delta g^{d}_{6}=-2.368(728)$ 
\\\cline{2-6}
& $g^e_{15}=0.069(97)$ 
& 
& 
& $\delta g^e_{15}=2.501(749)$
&  \\\cline{2-6}
&  $g^{a}_{3}=0.015(13)$
&  $g^{b}_{3}=0.02(24.13)$ 
&
& $\delta g^{a}_{3}=0.601(828)$
& $\delta g^{b}_{3}=2.805(724)$  
\\\cline{2-6}
\hline\hline
\end{tabular}
\end{table*}

\section{ Global analysis of charmed baryon two-body decays with strong phases}

Since the above analysis indicates that the matrix element of the penguin operator is likely to be on the order of $O(10^{-2})$, which is comparable to the matrix element of the current-current operator, significant CP violation (CPV) may occur. To study potential CP violation, it is essential to consider the complex form factor case.

Unfortunately, in the case of complex form factors, the 13 amplitudes in Eq.~\ref{su3} lead to 51 independent parameters, which cannot be effectively determined by the 35 available experimental data. Therefore, one has to reduce the number of amplitudes in the framework of SU(3).

Actually, according to the KPW theorem~\cite{Korner:1970xq,Pati:1970fg} the number of amplitudes in Eq.~\ref{su3} can be reduced.
To illustrate the flavor and color symmetries, the Hamiltonian can be decomposed into symmetric and antisymmetric components, using the current-current operator as an example,~\cite{He:2024pxh}
 \begin{eqnarray}
 \mathcal{H}_{eff}=\frac{G_F}{\sqrt{2}}\sum_{\lambda=\pm}C_{\lambda}(\mathcal{H}_\lambda)^{ij}_k [\bar q_i q^k]_{V-A}[\bar q_j c]_{V-A},\label{sh}
 \end{eqnarray}
where $C_\pm=(C_1\pm C_2)/2$ and  $(H_{+/-})^{ij}_k$ is  the symmetry/anti-symmetry for the indices $i,j$. 
The symmetry or antisymmetry of the anti-quark field also reflects the symmetry of its color.
Considering only the flavor structure, one can find that there are five different ways to contract the $SU(3)_f$ indices for $H_{15}$ in Eq.\ref{su3}: $a_{15}$, $b_{15}$, $c_{15}$, $d_{15}$, $e_{15}$. However, taking into account that color indices of quarks in $H_{15}$ (baryons) must be (anti)symmetric, these five terms can be consolidated into only one term, namely $e_{15}$~\cite{Geng:2023pkr}.
Similarly, for $H_{3}$, since only $a_3$ and $b_3$ contribute to the leading term, the aforementioned amplitudes can be further reduced as follows:
 \begin{eqnarray}\label{su3reduce}
\mathcal{M}^{IRA}&=&e_{15} \times(T_{c\bar{3}})_i(H_{15})^{jk}_l(\overline{T_8})^i_j P^l_k\notag\\
&+&a_{6} \times(T_{c\bar{3}})^{ik}(H_{\overline{6}})_{ij}
(\overline{T_8})^j_k P^l_l\notag\\
&+&b_{6} \times(T_{c\bar{3}})^{ik}(H_{\overline{6}})_{ij}(\overline{T_8})^l_k P^j_l\notag\\
&+&c_{6} \times(T_{c\bar{3}})^{ik}(H_{\overline{6}})_{ij}(\overline{T_8})^j_l P^l_k\notag\\
&+&d_{6} \times(T_{c\bar{3}})^{ik}(H_{\overline{6}})_{jl}(\overline{T_8})^j_i P^l_k.\notag\\
&+&a_{3} \times(T_{c\bar{3}})_{i}(H_{3})^{k}(\overline{T_8})^i_l P^l_k.\notag\\
&+&b_{3} \times(T_{c\bar{3}})_{i}(H_{3})^{k}(\overline{T_8})^i_k P^l_l.
\end{eqnarray}
 Reducing the number of decay amplitudes from 13 to 7 allows us to conduct a global analysis  on the available experimental data with complex form factors.

We also observe that the anti-symmetric part of the Hamiltonian in Eq.~(\ref{sh}) violates flavor symmetry, particularly in the $c\to d\bar{s}u$ processes. This also leads to the breaking of isospin symmetry.
Using the isospin-symmetric processes $(\Xi_c^0 \to p \pi^-, \Xi_c^+ \to n \pi^+)$ and $(\Xi_c^0 \to n \pi^0, \Xi_c^+ \to p \pi^0)$ as examples, the SU(3) amplitude $c_6$, corresponding to the anti-symmetric Hamiltonian, along with $c_{15}$ and $d_{15}$, corresponding to the symmetric part of the Hamiltonian, together break the isospin symmetry, as shown in Table.~\ref{tableamplitute}. As a result of this isospin symmetry violation, the predicted branching ratios and polarization parameters in Case I of Table.~\ref{dataeta} differ significantly.
After omitting the amplitudes $c_{15}$ and $d_{15}$ (Case II), one can derive the isospin symmetry relation as follows:
\begin{eqnarray}
&&\mathcal{M}(\Xi_c^0\to p\pi^-)=\mathcal{M}(\Xi_c^+\to n\pi^+)\notag\\
&&=-\sqrt{2}\mathcal{M}(\Xi_c^0\to n\pi^0)=\sqrt{2}\mathcal{M}(\Xi_c^+\to p\pi^0).
\end{eqnarray}
The situation also applies to 
 the isospin-symmetric processes $(\Xi_c^0 \to n\eta, \;\Xi_c^+ \to p\eta)$ and $(\Xi_c^0 \to n\eta^\prime, \Xi_c^+ \to p\eta^\prime)$. One can similarly derive their isospin symmetry relations after omitting the SU(3) amplitudes corresponding to the symmetric part of the Hamiltonian, as follows:
\begin{eqnarray}
&&\mathcal{M}(\Xi_c^0\to n\eta)=\mathcal{M}(\Xi_c^+\to p\eta),\notag\\
&&\mathcal{M}(\Xi_c^0\to n\eta^\prime)=\mathcal{M}(\Xi_c^+\to p\eta^\prime).
\end{eqnarray}
Therefore, in case II, the KPW theorem necessitates neglecting the color-suppressed SU(3) amplitudes $a_{15}, b_{15}, c_{15}, d_{15}$. This ensures that the isospin symmetry of the aforementioned decay channels will be naturally satisfied.

\begin{table*}[htbp!]
\caption{The experimental data~\cite{BESIII:2023uvs,Belle-II:2024jql,BESIII:2024sfz,PDG} and predicted values for branching ratios, polarization parameters ($\alpha,\beta,\gamma$) and CP violation with the final states $\eta^{(\prime)}$ for two different fits: real case (Case I in Eq.~\ref{su3}) and  complex case ( Case II means  in Eq.~\ref{su3reduce}). Here the data with $*$ mean the ratio of decay width between the corresponding processes and  $ \Gamma(\Xi_c^0\to \Xi^-\pi^+)$. }\label{dataeta}
\begin{tabular}{|c|c|c|c|c|c|c|c|c|c|c}\hline\hline
\multirow{2}{*}{channel} &  \multicolumn{2}{c|}{ exp}&  \multicolumn{2}{c|}{ prediction (Case I)}
&\multicolumn{5}{c|}{ prediction (Case II)}
\cr\cline{2-10}   
& Br($\%$)& $\alpha$ 
& Br($\%$)& $\alpha$ 
&  Br($\%$)& $\alpha$ & $\beta$&$\gamma$&$A_{CP}$
 \\\hline \hline
$\Lambda^{+}_{c}\to \Sigma^{+}  \eta $ & $0.32(5)$
&$-0.99(6)$&$0.318(48)$ &$-0.999(33)$&0.302(50)&-0.962(57)&-0.26(23)&-0.10(33)& \\\hline
$\Lambda^{+}_{c}\to \Sigma^{+}  \eta^\prime $ & $0.41(8)$
&$-0.460(67)$& $0.409(79)$&$-0.459(67)$&0.271(65)&-0.452(67)&-0.6(1.1)&-0.69(94)&\\\hline
$\Lambda^{+}_{c}\to p  \eta $  &$0.158(11)$ &    &$0.158(11)$ &-0.71(12)&0.158(11) &0.3(185.3)&0.6(456.0)&-0.7(381.8)&-0.2(85.4)\\\hline
$\Lambda^{+}_{c}\to p  \eta^\prime $        & $0.0484(91)$
&  &$0.0484(91)$ &0.89(38)&0.0484(91) &0.2(1900)&0.7(1089.3)&0.7(944.1)&-0.5(339.9)\\\hline
$\Xi^{+}_{c}\to \Sigma^{+} \eta $   & & & $0.356(66)$&$-0.22(20)$&0.1(38.0)& 0.7(311.4)&-0.06(536.28)&0.8(229.5)&-0.04(349.28) \\\hline
$\Xi^{+}_{c}\to \Sigma^{+}  \eta^\prime $   & & & $0.87(18)$& $-0.18(18)$&0.4(226.6)&-0.9(467.2)&-0.4(995.1)&-0.1(52.7)&0.4(205.3)\\\hline
$\Xi^{+}_{c}\to p  \eta $ &&		&$0.0202(16)$ &$-0.731(59)$&0.0172(22)&-0.54(13)&0.15(35)&-0.828(88)& \\\hline
$\Xi^{+}_{c}\to p  \eta^\prime $ 		&&&$0.00163(53)$ &$0.18(35)$&0.0096(29)&-0.14(11)&-0.15(72)&-0.98(11)& \\\hline
$\Xi^{0}_{c}\to \Xi^{0}  \eta $ & $0.110(14)^*$
& &  $0.079(19)$&0.95(24)& 0.330(41)&0.52(14)&0.85(12)&0.10(69)&\\\hline
$\Xi^{0}_{c}\to \Xi^{0}  \eta^\prime $ & $0.080(22)^*$
& &$0.084(28)$ &-0.82(77)&0.335(57)&-0.775(81)&-0.2(1.1)&-0.60(42)&\\\hline
$\Xi^{0}_{c}\to \Sigma^{0}  \eta $    && &$0.046(10)$&$0.13(20)$&0.02(6.39)& 0.7(311.2)&-0.06(535.14)&0.8(298.2)&-0.03(350) \\\hline
$\Xi^{0}_{c}\to \Sigma^{0}  \eta^\prime$    &&&$0.086(30)$ &$-0.50(21)$&0.07(37.85)& -0.9(467.3)&-0.4(995.7)&-0.1(52.7)&0.4(204.2) \\\hline
$\Xi^{0}_{c}\to \Lambda  \eta $  &&  & $0.0411(90)$&$-0.9992(84)$& 0.05(8.80)&0.4(99.1)&-0.4(321.3)&-0.8(111.2)&-0.07(156.00) \\\hline
$\Xi^{0}_{c}\to \Lambda  \eta^\prime$  &&  & $0.030(23)$&$0.72(48)$&0.03(11.81)&0.2(815.8)&0.3(1216.0)&-0.9(230.3)&-0.4(313.0) \\\hline
$\Xi^{0}_{c}\to n \eta $  &&     &$0.00222(68)$ &$-0.94(11)$& 0.00576(75) &-0.54(13)&0.15(35)&-0.828(88)&\\\hline
$\Xi^{0}_{c}\to n \eta^\prime $  &&     & $0.00063(44)$&$0.19(77)$&0.00323(97) &-0.14(11)&-0.15(72)&-0.98(11)&\\\hline
\hline
\end{tabular}
\end{table*}

Considering the 7 SU(3) amplitude, the 14 complex form factors can be expressed as
 \begin{eqnarray}
&&|A^e_{15}|e^{i\delta^e_{15}},\quad|A^a_{6}|e^{i\delta^a_{6}},\quad|A^b_{6}|e^{i\delta^b_{6}},\quad|A^c_{6}|e^{i\delta^c_{6}},\notag\\
&&|A^d_{6}|e^{i\delta^d_{6}},\quad|A^a_{3}|e^{i\delta^a_{3}},\quad|A^b_{3}|e^{i\delta^a_{3}}, \quad A=f,g,
 \end{eqnarray}
where each phase of the form factors incorporates both the weak and strong phases. Actually, the strong phases and weak phases come from the different sources.
The strong phases arise from potential contributions of intermediate on-shell states in the decay process~\cite{PDG}, induced by rescattering processes and loop effects. This feature is crucial for observing CP violation and understanding weak interactions.
On the other hand, weak phases primarily originate from CKM matrix elements $\lambda = V_{cq}^*V_{uq'}$ and $\lambda_b = V_{cb}^*V_{ub}$ in the weak interaction Hamiltonian, as shown in Eq.~\ref{lagrangian}. Thus, when the Hamiltonian is determined, the weak phase is concurrently determined. 
Since the weak phases are not explicitly provided in the SU(3) matrix elements of the Hamiltonian, we can extract them from the form factors, which are expressed as the combination of the absolute values of weak and strong phases
 \begin{eqnarray}
 &&|A^e_{15}|e^{i\delta A^e_{15}}e^{i\phi_1},|A^a_{6}|e^{i\delta A^a_{6}}e^{i\phi_1},|A^b_{6}|e^{i\delta A^b_{6}}e^{i\phi_1},\notag\\
 &&|A^c_{6}|e^{i\delta A^c_{6}}e^{i\phi_1},
 |A^d_{6}|e^{i\delta A^d_{6}}e^{i\phi_1},|A^a_{3}|e^{i\delta A^a_{3}}e^{i\phi_T},\notag\\
 &&|A^b_{3}|e^{i\delta A^a_{3}}e^{i\phi_T}, A=f,g,
\end{eqnarray}
where the weak phases $\phi_1$ and $\phi_T$ arise from the current-current operator and penguin operator with $\phi_1=\arg(V_{cq}^*V_{uq'})\approx 0,-\pi$ and $\phi_T=\arg(V_{cb}^*V_{ub})\approx 1.147$, respectively.

Factorizing out the global phase and set it to zero $\delta f_6^a = 0$, we now have 27 parameters to fit the 35 experimental data, which enables a comprehensive global fit. 
The fit results (Case II) are presented in Table.\ref{tablefit}. 
Based on the fitted form factors, we predict the corresponding observables as shown in Table.~\ref{dataeta} and \ref{data}. 
During our analysis, we observed that the $Br(\Xi_c^0 \to \Xi^-\pi^+)$ contributes significantly to the $\chi^2$. Excluding this data reduces the $\chi^2/d.o.f$ from 5.95 to 1.19, which is consistent with other studies~\cite{Geng:2023pkr}.
However, there are seven experimental data measured by the ratio $Br(\Xi_c^0 \to BP)/Br(\Xi_c^0 \to \Xi^-\pi^+)$, making this channel crucial and difficult to be excluded. Our prediction for $Br(\Xi_c^0 \to \Xi^-\pi^+)$ is $(3.10 \pm 0.12)\%$. 
Considering a similar discrepancy observed in $Br(\Xi_c^0 \to \Xi^-e^+\nu_e)$, we suspect reasonably that the current experimental value for $Br(\Xi_c^0 \to \Xi^-\pi^+)$ might underestimate its true value, necessitating further scrutiny~\cite{Geng:2023pkr}.
Similarly, for rigor in our research, we acknowledge the possibility that the leading amplitude in the KPW theorem may not perfectly explain CBTD. Therefore, we eagerly await future experimental data to confirm  our conclusions.

\begin{table*}[htbp!]
\caption{
The experimental data and predicted values for branching ratios, polarization parameters ($\alpha,\beta,\gamma$) and CP violation
for two different fits. 
The details are the same with Table.~\ref{dataeta}.
}\label{data}\begin{tabular}{|c|c|c|c|c|c|c|c|c|c|c}\hline\hline
\multirow{2}{*}{channel} &  \multicolumn{2}{c|}{ exp}&  \multicolumn{2}{c|}{ prediction (Case I)} 
&  \multicolumn{5}{c|}{prediction (Case II)}\cr\cline{2-10}   & 
Br($\%$)& $\alpha$ & 
Br($\%$)& $\alpha$ &   Br($\%$)& $\alpha$ & $\beta$ & $\gamma$ & $A_{CP}$
 \\\hline \hline
$\Lambda^{+}_{c}\to \Sigma^{0}  \pi^{+} $ & $1.27(6)$
&$-0.466(18)$&$1.264(49)$ &$-0.473(15)$ & 1.220(47)&-0.470(15)&0.73(33)&-0.50(48)&\\\hline
$\Lambda^{+}_{c}\to \Lambda  \pi^{+} $ 
& $1.29(5)$
&$-0.755(6)$& $1.274(48)$&$-0.7551(60)$&1.259(48)&-0.7545(60)&0.12(50)&-0.645(93)&\\\hline
$\Lambda^{+}_{c}\to \Sigma^{+}  \pi^{0} $ & $1.24(9)$
&$-0.484(27)$&$1.278(50)$ &$-0.471(15)$& 1.231(48)& -0.469(15)&0.72(33)&-0.51(48)&\\\hline
$\Lambda^{+}_{c}\to p  K_{S}^{0} $ & 1.59(7) 
& 0.2(5)&$1.621(49)$ &$-0.897(88)$&1.579(70)&0.03(50)&0.2(1.5)&-0.98(34)&\\\hline
$\Lambda^{+}_{c}\to \Xi^{0}  K^{+} $ &$0.55(7)$
&0.01(16)&$0.544(52)$ &$-0.067(89)$&0.455(39)&-0.05(16)&0.15(59)&0.988(87)& \\\hline
$\Xi^{+}_{c}\to \Sigma^{+}  K_{S}^{0} $ &&&$1.10(35)$ &$0.939(53)$&0.7(2.0)&-0.2(2.4)&0.89(71)&0.4(1.5)&\\\hline
$\Xi^{+}_{c}\to \Xi^{0}  \pi^{+} $ & $1.6(8)$
&&$0.71(12)$ &0.966(33)&1.15(55)&-0.17(54)&0.97(21)&-0.19(80)&\\\hline
$\Xi^{0}_{c}\to \Sigma^{0}  K_{S}^{0} $ &$0.0380(70)^*$
&&$0.055(14)$ & 0.95(18)  &0.120(21)&0.6(1.1)&0.6(2.4)&0.5(2.2)&\\\hline
$\Xi^{0}_{c}\to \Lambda  K^0_S$  & $0.225(13)^*$
& &$0.324(60)$ & -0.59(32)&0.724(27) &-0.27(32)&0.48(83)&-0.83(46)&\\\hline
$\Xi^{0}_{c}\to \Sigma^{+}  K^{-} $ &$0.123(12)^*$
& &$0.177(36)$ &0.57(57)&0.406(33)&-0.07(22)&0.20(81)&0.98(17)& \\\hline
$\Xi^{0}_{c}\to \Xi^{-}  \pi^{+} $ & $1.43(27)$
&$-0.640(51)$& $1.44(26)$&$-0.635(49)$&3.01(12) & -0.689(34) &0.37(44)&-0.62(24)&\\\hline
$\Xi^{0}_{c}\to \Xi^{0}  \pi^{0} $ & $0.480(36)^*$ 
& $-0.90(28)$  &$0.7(12)$&$-0.9982(69)$  &0.752(60)&-0.72(13)&-0.08(47)&-0.69(17)&\\\hline\hline
$\Lambda^{+}_{c}\to \Sigma^{0}  K^{+} $  & $0.0370(31)$
&$-0.54(20)$ &$0.0365(30)$ &$-0.58(18)$& 0.0368(28)&-0.75(13)&-0.09(50)&-0.66(18)&\\\hline
$\Lambda^{+}_{c}\to \Lambda  K^{+} $  & $0.0642(31)$
&$-0.58(5)$ &$0.0647(30)$ &$-0.575(48)$&0.0637(31)&-0.572(50)&0.96(16)&-0.12(95)&0(0)\\\hline
$\Lambda^{+}_{c}\to \Sigma^{+} K^0_{S,L} $  & $0.047(14)$
& &$0.0494(78)$& -0.58(24) &0.0372(29)&-0.75(13)&-0.09(49)&-0.66(18)& \\\hline
$\Lambda^{+}_{c}\to p  \pi^{0} $       
& 0.0156(75)
&
&$0.0196(62)$ &0.71(24)&0.0161(75)& 0.85(71) &-0.07(4.44)&0.51(87)&-0.38(44)\\\hline
$\Lambda^{+}_{c}\to n  \pi^{+}$            &  $0.066(13)$
&&$0.073(11)$ & 0.12(13) &0.070(12)& -0.56(57)&0.05(86)&-0.83(36)&0.17(37) \\\hline
$\Xi^{+}_{c}\to \Sigma^{0}  \pi^{+} $      & & &$1.725(98)$&$-0.679(35)$&0.45(10)&-0.17(28)&0.83(35)&-0.52(55)&0.06(10) \\\hline
$\Xi^{+}_{c}\to \Lambda  \pi^{+} $    &&&$0.808(54)$ &$-0.899(40)$&0.056(39)&0.34(41)&-0.83(50)&0.44(92)&0.18(46)\\\hline
$\Xi^{+}_{c}\to \Sigma^{+}  \pi^{0} $  && &$1.421(76)$ &$-0.978(17)$&0.31(14)&0.17(57)&0.87(70)&-0.5(1.2)&0.008(137)   \\\hline
$\Xi^{+}_{c}\to p  K^0_{S,L} $          &&&$0.137(18)$ &$-0.78(25)$&0.190(19)&-0.55(13)&-0.06(36)&-0.834(97) & \\\hline
$\Xi^{+}_{c}\to \Xi^{0}  K^{+} $         & &&$0.078(15)$&$0.31(16)$&0.060(37)&-0.31(51)&-0.5(1.1)&0.84(50)&0.01(44)   \\\hline
$\Xi^{0}_{c}\to \Sigma^{0}  \pi^{0} $    & &&$0.0315(73)$&$-0.89(12)$& 0.033(20)&0.13(87)&-0.4(1.5)&-0.91(52)&-0.08(20)  \\\hline
$\Xi^{0}_{c}\to \Lambda  \pi^{0} $    && &$0.0486(61)$ &$-0.81(12)$&0.016(21) &-0.76(48)&-0.43(83)&0.49(59)&-0.008(349)\\\hline
$\Xi^{0}_{c}\to \Sigma^{+}  \pi^{-} $     &&&$0.401(46)$&$-0.929(38)$&0.0231(19)&-0.07(24)&0.21(86)&0.97(19)& \\\hline
$\Xi^{0}_{c}\to p  K^{-} $                &&&$0.618(58)$ &$-0.926(34)$& 0.0207(23)&-0.09(29)&0.3(1.0)&0.96(28)&\\\hline
$\Xi^{0}_{c}\to \Sigma^{-}  \pi^{+} $     &&&$0.171(26)$ &$0.004(102)$&0.244(34) &-0.42(25)&0.48(52)&-0.77(32)&0.04(14)\\\hline
$\Xi^{0}_{c}\to n  K^0_{S,L} $& &&$0.0197(86)$&$-0.95(16)$ &0.076(10)&-0.0387(51)&0.60(35)&-0.70(32)&\\\hline
$\Xi^{0}_{c}\to \Xi^{-}  K^{+} $          & $0.0275(57)^*$
&&$0.0473(98)$ &-0.962(46) &0.102(14)&-0.90(17)&0.10(44)&-0.43(39)&0.06(22)\\\hline
$\Xi^{0}_{c}\to \Xi^{0}  K^0_{S,L} $  &&&$0.0155(45)$ &$-0.78(27)$&0.0381(26) &-0.496(33)&0.77(30)&1(0)&\\\hline\hline
$\Lambda^{+}_{c}\to p  K^0_L $ 	 &  1.67(7)
& & $1.637(52)$& -0.84(11)& 1.650(69)  & 0.08(55)&0.3(1.6)&-0.96(44) & \\\hline
$\Lambda^{+}_{c}\to n  K^{+} $ 		&&& 0.00086(15)&$0.99999(59)$&0.00157(70)&-0.16(49)&0.89(35)&-0.43(68)& \\\hline
$\Xi^{+}_{c}\to \Sigma^{0}  K^{+} $ & &&$0.01311(37)$&$-0.6717(67)$& 0.01207(46)&-0.657(28)&0.35(43)&-0.67(22)&\\\hline
$\Xi^{+}_{c}\to \Lambda  K^{+} $&&&$0.00169(14)$ &$-0.012(38)$&0.00220(43) &-0.34(18)&0.938(85)&-0.06(80)&\\\hline
$\Xi^{+}_{c}\to \Sigma^{+}  K^0_L $ & & &$1.20(36)$&$0.99998(91)$&0.9(2.2)&0.03(2.83)&0.96(45)&0.3(1.6) & \\\hline
$\Xi^{+}_{c}\to p  \pi^{0} $ 		&& &$0.00116(68)$&$0.16(77)$& 0.00172(20)&-0.09(30) &0.3(1.1)&0.96(31)&\\\hline
$\Xi^{+}_{c}\to n  \pi^{+} $ 		&&&$0.00407(38)$ &$-0.13(17)$&0.00344(40)& -0.09(30)&0.3(1.1)&0.96(31)&\\\hline
$\Xi^{0}_{c}\to \Sigma^{0}  K_{L}^{0}$ &&&$0.079(17)$ &0.82(26)&0.131(23) &0.8(1.0)&0.5(3.0)&0.4(2.2)&\\\hline
$\Xi^{0}_{c}\to \Lambda  K^0_L$  && &$0.308(53)$&$-0.53(32)$&0.653(22)&-0.23(37) &0.45(95)&-0.87(47)&\\\hline
$\Xi^{0}_{c}\to p  \pi^{-} $ 		&& &$0.00091(18)$ &$0.43(47)$& 0.00116(14)&-0.09(30)&0.3(1.1)&0.96(31)&\\\hline
$\Xi^{0}_{c}\to \Sigma^{-}  K^{+} $ &&&$0.00389(69)$&$-0.601(48)$& 0.00806(31)&-0.658(28)&0.35(43)&-0.67(22)&\\\hline
$\Xi^{0}_{c}\to n  \pi^{0} $ 		&&&$0.00058(44)$&$-0.14(22)$&0.000578(68)&-0.09(30)&0.3(1.1)&0.96(31)& \\\hline
\hline
\end{tabular}
\end{table*}


In terms of the amplitude corresponding to $H_3$, the $g^a_3$ from global fit is  nonzero and the error is small. On the other hand, although the central values of $f^b_3$ and $g^b_3$ are not small, the large errors make it impossible to estimate the central value precisely.
Additionally, it is noted that predictions for some processes have relatively large errors, such as $\Lambda_c^+ \to p \eta^{(\prime)}$, $\Xi_c^+ \to \Sigma^+ \eta^{(\prime)}$, due to form factors with large errors.
Therefore, a comprehensive analysis with sufficient experimental data is necessary in the future.
Nevertheless, the better fit results judged by $\chi^2/d.o.f$ may suggest that the results from Case I are reasonable.

When considering the weak and strong phases, CP violation becomes the most intriguing issue. 
CPV requires two distinct types of weak phases from the current-current operator and penguin diagrams involving the CKM matrix, as shown in Eq.\ref{lagrangian}. 
Using the fitted results (Case II) from Table.~\ref{tablefit}, one can predict the direct CPV ($A_{CP}$) for Cabibbo-suppressed processes:
\begin{eqnarray}
A_{CP}(B_c\to B M)=\frac{\Gamma(B_c\to B M)-\Gamma(\bar B_c\to \bar B \bar M)}{\Gamma(B_c\to B M)+\Gamma(\bar B_c\to \bar B \bar M)}.
\end{eqnarray}
The corresponding CPV for Cabibbo-suppressed processes is provided in the last column of Table.~\ref{dataeta} and \ref{data}. Unfortunately, due to the large error associated with the contribution of $H_3$ in our fit, including the errors the predicted CPV is also approximately zero.

Hence, our study suggests that the contribution of the $H_3$ amplitudes induced by penguin operator is indeed significantly. However, the current experimental data are insufficient for fully determining the contribution of $H_3$ and accurately predicting CP violation (CPV) through global analysis. This underscores the necessity for additional experimental measurements to precisely determine the role of $H_3$ in future studies.\\

\section{Conclusion}

In our study, we investigate the charmed baryon two-body decay processes using SU(3) flavor symmetry.
By including the penguin operator ignored in previous studies, we add a new Hamiltonian matrix $H_3$, which increases the number of amplitudes in the IRA method from 9 to 13.
Along with a significant amount of measured experimental data, we consider the previously ignored contribution from the penguin diagram to perform a global analysis.

For real form factors (Case I), we obtain $\chi^{2}/d.o.f = 0.788$, indicating that SU(3) flavor symmetry, including the penguin contribution, explains the measured data exceptionally well.
Additionally, we predict some unmeasured experimental observables in Tables.~\ref{dataeta} and \ref{data}, which can be further verified in future measurements.
Based on the global fit results we obtained in Table.~\ref{tablefit}, we can significantly estimate the contribution of amplitudes and found that the contribution of the $H_3$ amplitude in charmed baryon two-body decay processes is of the order $\sim 0.01$, which is comparable to the contribution of the tree-level diagram.
This indicates that for the CKM matrix with an order of $O(10^{-4})$, the strong interaction and other long-distance interactions could make the effects of the hadronic matrix element large.

In order to extend the complex form factors for studying possible CP violation, we use the KPW theorem to reduce the SU(3) invariant amplitudes to seven.
However, the analysis with the complex form factor significantly conflicts with the experimental data $Br(\Xi_c^0\to\Xi^-\pi^+)$, and by excluding this data, $\chi^2/d.o.f$ is reduced from $5.95$ to $1.19$.
Although the analysis with complex form factors shows a significant central value of the $H_3$ amplitude contribution, the large error from the corresponding form factors makes it challenging to precisely determine its true contribution.
Consequently, the direct CP violation in decay processes is predicted to be approximately zero.
With more data in future experiments, the penguin diagram contribution with the amplitude proportional to $V_{cb}^*V_{ub}$ will be precisely determined, allowing for a more accurate prediction of CP violation.
Our work encourages further theoretical investigations and experimental measurements in the future.

\section*{Acknowledgements}
The work of Zhi-Peng Xing is supported by NSFC under grant No.12375088 and No. 12335003. The work of Yu Ji Shi is supported by Opening Foundation of Shanghai Key Laboratory of Particle Physics and Cosmology under Grant No.22DZ2229013-2 and National Natural Science Foundation of China under Grant No.12305103. The work of Jin Sun is supported by IBS under the project code, IBS-R018-D1.  The work of Ye Xing is supported by NSFC under grant No.12005294.

\appendix
\section{ The  decomposition of IRA heavy-to-light Hamiltonian }
 
There is another form of decomposition of the IRA heavy-to-light Hamiltonian as
 \begin{eqnarray}
H^{ij}_k&=&\frac{1}{8}(H_{15})^{ij}_k+\frac{1}{4}\epsilon^{ijl}(H_{\bar 6})_{lk}+\delta^j_k\left(\frac{3}{8}(H^t_3)^i-\frac{1}{8}(H^p_3)^i\right)\notag\\
&&+\delta^i_k\left(\frac{3}{8}(H^p_3)^j-\frac{1}{8}(H^t_3)^j\right),\label{irahb}
 \end{eqnarray}
 where $(H^t_3)^i=H^{im}_m$ and $(H^p_3)^i=H^{mi}_m$.
At first glance, we might naively say that the expression of the decomposition is very different from our formula in Eq.~\ref{irah}. Actually, the two expressions are equivalent, as can be clearly understood from the operator aspect of the IRA Hamiltonian.
 In this aspect, the Hamiltonian matrix   $H^{ij}_k$ is expressed as
  \begin{eqnarray}
O(H)^{ij}_k=[\bar q_j c][\bar q_i q_k], \quad (q_1,q_2,q_3)=(u,d,s),
 \end{eqnarray}
where we omit the color and Lorentz structure in SU(3) symmetry. 
Correspondingly, the matrix elements of $H_{15}$, $H_{\bar 6}$, and $H_3$ both can be  obtained. 
By using the transformation relations of $H_{3/{3\prime}}$ and $H_3^{t/p}$ 
\begin{eqnarray}
(H_3)^{ij}_k&=&\delta^i_k (H^t_3)^j+\delta^j_k (H^p_3)^i,\notag\\(H_{3\prime})^{ij}_k&=&\delta^i_k (H^p_3)^j+\delta^j_k (H^t_3)^i,
 \end{eqnarray}
one can directly find that the $3$ representations in Eq.~\ref{irah} and Eq.~\ref{irahb} are equivalent. And the  the matrix elements $H_3^{t/p}$ are 
  \begin{eqnarray}
O(H_3^p)^1&=&[\bar u c][\bar u u]+[\bar u c][\bar d d]+[\bar u c][\bar s s],\notag\\
O(H_3^p)^2&=&[\bar d c][\bar u u]+[\bar d c][\bar d d]+[\bar d c][\bar s s],\notag\\
O(H_3^p)^3&=&[\bar s c][\bar u u]+[\bar s c][\bar d d]+[\bar s c][\bar s s],\notag\\
O(H_3^t)^1&=&[\bar u c][\bar u u]+[\bar d c][\bar u d]+[\bar s c][\bar u s],\notag\\
O(H_3^t)^2&=&[\bar u c][\bar d u]+[\bar d c][\bar d d]+[\bar s c][\bar d s],\notag\\
O(H_3^t)^3&=&[\bar u c][\bar s u]+[\bar d c][\bar s d]+[\bar s c][\bar s s].
 \end{eqnarray}
 Besides, the matrix elements of $H_{\bar6}$ and $H_{15}$ are
 \begin{widetext}
 \begin{eqnarray}
O(H_{\bar 6})^{11}&=&[\bar d c][\bar s u]-[\bar s c][\bar d u],\;\;O(H_{\bar 6})^{22}=[\bar s c][\bar u d]-[\bar u c][\bar s d],\;\;O(H_{\bar 6})^{33}=[\bar u  c][\bar d s]-[\bar d c][\bar u s],\notag\\
O(H_{\bar 6})^{12}&=&\frac{1}{2}([\bar s c][\bar u u]-[\bar u c][\bar s u]+[\bar d c][\bar s d]-[\bar s c][\bar d d]),\;\;O(H_{\bar 6})^{23}=\frac{1}{2}([\bar u c][\bar d d]-[\bar d c][\bar u d]+[\bar s c][\bar u s]-[\bar u c][\bar ss ]),\notag\\
O(H_{\bar 6})^{31}&=&\frac{1}{2}([\bar d c][\bar s s]-[\bar s c][\bar d s]+[\bar u c][\bar d u]-[\bar d c][\bar u u]),\notag\\
O(H_{15})^{11}_1&=&[\bar u c][\bar u u]-\frac{1}{2}([\bar u c][\bar d d]+[\bar d c][\bar u d]+[\bar s c][\bar u s]+[\bar u c][\bar ss ]),\quad  O(H_{15})^{23}_1=[\bar s c][\bar d u ]+[\bar d c][\bar s u ]\notag\\
O(H_{15})^{22}_2&=&[\bar d c][\bar d d]-\frac{1}{2}([\bar u c][\bar d u]+[\bar s c][\bar d s]+[\bar d c][\bar u u]+[\bar d c][\bar ss ]),\quad  O(H_{15})^{13}_2=[\bar u c][\bar s d  ]+[\bar s c][\bar u d ]\notag\\
O(H_{15})^{33}_3&=&[\bar s c][\bar s s]-\frac{1}{2}([\bar u c][\bar s u]+[\bar d c][\bar d s]+[\bar s c][\bar u u]+[\bar s c][\bar dd ]),\quad  O(H_{15})^{12}_3=[\bar d c][\bar u s  ]+[\bar u c][\bar d s ]\notag\\
O(H_{15})^{21}_1&=&\frac{3}{4}([\bar d c][\bar u u]+[\bar u c][\bar d u])-\frac{1}{2}[\bar d c][\bar d d]-\frac{1}{4}([\bar s c][\bar d s]+[\bar d c][\bar s s]),\;O(H_{15})^{11}_2=2[\bar u c][\bar u d],\notag\\
O(H_{15})^{21}_2&=&\frac{3}{4}([\bar d c][\bar u d]+[\bar u c][\bar d d])-\frac{1}{2}[\bar u c][\bar u u]-\frac{1}{4}([\bar s c][\bar u s]+[\bar u c][\bar s s]),\;O(H_{15})^{11}_3=2[\bar u c][\bar u s],\notag\\
O(H_{15})^{31}_1&=&\frac{3}{4}([\bar s c][\bar u u]+[\bar u c][\bar s u])-\frac{1}{2}[\bar s c][\bar s s]-\frac{1}{4}([\bar d c][\bar  s d]+[\bar s c][\bar d d]),\;O(H_{15})^{22}_3=2[\bar d c][\bar d s],\notag\\
O(H_{15})^{31}_3&=&\frac{3}{4}([\bar s c][\bar u s]+[\bar u c][\bar s s])-\frac{1}{2}[\bar u c][\bar u u]-\frac{1}{4}([\bar d c][\bar  u d]+[\bar u c][\bar d d]),\;O(H_{15})^{22}_1=2[\bar d c][\bar d u],\notag\\
O(H_{15})^{32}_2&=&\frac{3}{4}([\bar s c][\bar dd]+[\bar d c][\bar s d])-\frac{1}{2}[\bar s c][\bar ss]-\frac{1}{4}([\bar u c][\bar su]+[\bar s c][\bar uu]),\;O(H_{15})^{33}_1=2[\bar s c][\bar s u],\notag\\
O(H_{15})^{32}_3&=&\frac{3}{4}([\bar s c][\bar ds]+[\bar d c][\bar s s])-\frac{1}{2}[\bar d c][\bar dd]-\frac{1}{4}([\bar u c][\bar du]+[\bar d c][\bar uu]),\;O(H_{15})^{33}_2=2[\bar s c][\bar  s d].
 \end{eqnarray}
  \end{widetext}
Comparing with the previous work~\cite{He:2000ys, Wang:2020gmn}, one can find that the expressions of $H_{\bar 6}$ and $H_{15}$ in Eq.~\ref{irah} and Eq.~\ref{irahb} can be transformed by the factors $-\frac{1}{2}$ and $\frac{1}{4}$, respectively. These two factors  exactly compensate for the difference in coefficients of Hamiltonian.
Therefore, after considering the differences from both Hamiltonian and  the operators, the decomposition formulas in Eq.~\ref{irah} and Eq.~\ref{irahb} are equivalent to each other.

\bibliographystyle{JHEP}
\bibliography{ref}

\end{document}